\documentclass[%
 reprint,
superscriptaddress,
 amsmath,amssymb,
aps,
]{revtex4-2}

\usepackage{graphicx}
\usepackage{dcolumn}
\usepackage{bm}
\usepackage[dvipsnames]{xcolor}
\usepackage{amsmath}
\usepackage{mathtools}
\usepackage{braket}

\DeclareUnicodeCharacter{2212}{-}

\newcommand{\Ext}{\mathsf{Ext}}
\newcommand{\esound}{\varepsilon_{\mathrm{sou}}}
\newcommand{\ecom}{\varepsilon_{\mathrm{com}}}
\newcommand{\esmooth}{\epsilon_{s}}
\newcommand{\eExt}{\epsilon_{\mathrm{ext}}}
\newcommand{\eEAT}{\epsilon_{\mathrm{EA}}}
\newcommand{\vb}[1]{\mathbf{#1}}
\newcommand{\norm}[1]{\left \lVert #1 \right \rVert}
\newcommand{\Tr}{\mathrm{Tr}}
\newcommand{\floor}[1]{\left \lfloor #1 \right \rfloor}
\newcommand{\vbf}[1]{\boldsymbol{#1}}
\newcommand{\pg}{P_{\mathrm{guess}}}
\begin{document}

\preprint{APS/123-QED}

\title{Self-testing quantum randomness expansion on an integrated photonic chip}

\author{Gong Zhang}
 \email{zhanggong@nus.edu.sg}
\affiliation{Department of Electrical \& Computer Engineering, National University of Singapore, Singapore}

\author{Ignatius William Primaatmaja}
\affiliation{Department of Electrical \& Computer Engineering, National University of Singapore, Singapore}
\affiliation{Squareroot8 Technologies Pte Ltd, Singapore}
 \affiliation{Global Technology Applied Research, JPMorganChase, New York, NY 10017 USA}
\author{Yue Chen}
\affiliation{Department of Electrical \& Computer Engineering, National University of Singapore, Singapore}
\author{Si Qi Ng}
\affiliation{Department of Electrical \& Computer Engineering, National University of Singapore, Singapore}
\affiliation{Centre for Quantum Technologies, National University of Singapore, Singapore}
\author{Hong Jie Ng}
\affiliation{Department of Electrical \& Computer Engineering, National University of Singapore, Singapore}
\affiliation{Squareroot8 Technologies Pte Ltd, Singapore}
 \author{Marco Pistoia}
 \affiliation{Global Technology Applied Research, JPMorganChase, New York, NY 10017 USA}
\author{Xiao Gong}
\affiliation{Department of Electrical \& Computer Engineering, National University of Singapore, Singapore}
\author{Koon Tong Goh}
\affiliation{Squareroot8 Technologies Pte Ltd, Singapore}
\author{Chao Wang}
\affiliation{Department of Electrical \& Computer Engineering, National University of Singapore, Singapore}

\author{Charles Lim}
\email{charleslim.research@gmail.com}
\affiliation{Department of Electrical \& Computer Engineering, National University of Singapore, Singapore}
 \affiliation{Global Technology Applied Research, JPMorganChase, New York, NY 10017 USA}
\affiliation{Centre for Quantum Technologies, National University of Singapore, Singapore}

\date{\today}
\begin{abstract} 
The power of quantum random number generation is more than just the ability to create truly random numbers---it can also enable \emph{self-testing}, which allows the user to verify the implementation integrity of certain critical quantum components with minimal assumptions. In this work, we develop and implement a self-testing quantum random number generator (QRNG) chipset capable of generating 15.33 Mbits of certifiable randomness in each run (an expansion rate of $5.11\times 10^{-4}$ at a repetition rate of 10 Mhz). The chip design is based on a highly loss-and-noise tolerant measurement-device-independent protocol, where random coherent states encoded using quadrature phase shift keying (QPSK) are used to self-test the quantum homodyne detection unit---well-known to be challenging to characterise in practice. Importantly, this proposal opens up the possibility to implement miniaturised self-testing QRNG devices at production scale using standard silicon photonics foundry platforms.

\end{abstract}

\maketitle


\section{\label{sec:intro}Introduction}

Random number generation (RNG) is important for a multitude  of reasons, particulary for computing, cryptography, and gaming. Roughly speaking, the quality of random numbers can be evaluated using two metrics: uniformity and unpredictability~\cite{SP800_22}. The former looks at how much entropy the RNG device is able to produce---the higher the entropy, the more unbiased the output is. The latter studies the amount of residual entropy when the observer is given some additional information about the device. In the ideal setting, the device should have maximum entropy even when the observer is given full information about the device.

True random number generators (TRNGs) provide a practical solution to both uniformity and unpredictability by leveraging random classical physical phenomena like shot noise and brownian motion~\cite{Stipčević2014, yao2020thermal}. To estimate the entropy of the TRNG device, a careful characterisation of the underlying physical phenomenon is needed; normally done at the start of the product lifecycle. However, this approach can be unreliable as physical phenomena can experience changes over the time, or if the mathematical model is unable to fully describe the phenomenon for the purpose of entropy estimation (in this case, the increased possibility of an zero-day exploit).

A promising solution is to use quantum random number generators (QRNGs)~\cite{herrero2017quantum,ma2016quantum,liu2024post}. These devices utilize the inherent unpredictability of quantum systems to provide high-quality random bits. However, while QRNGs hold great promise for generating truly unpredictable random bits, the security of today's commercial QRNGs is still dependent on the accurate modeling and characterization of the system components. 

In order to achieve a higher standard of security, it is necessary to allow users to self-test the device’s integrity against side information and component degradation without relying on precise device specifications~\cite{mayers2004self}. Additionally, in order to provide a miniaturized and cost-effective QRNG, chip integration using relatively mature photonic platforms is essential. However, achieving a high-quality, self-testing QRNG implementation on a chip poses significant challenges for both the QRNG protocol and the experimental scheme. 

Across different self-testing QRNG protocols, there exists a trade-off between security and practicality, with varying assumptions about the devices. Device-independent (DI) QRNGs~\cite{acin2016certified,pironio2010random,liu2021device,shalm2021device,deng2013exploring,Vivoli_2015}, which rely only on minimum physical assumptions and are independent of the physical implementation, face limitations during implementation due to the requirement of high-efficiency single-photon detectors and entangled photon sources, which restrict the system footprint and scalability. 

Semi-DI-QRNGs improve feasibility by introducing practical assumptions. 
One such assumption is that the source states can be easily well-characterized while maintaining the independence of measurement devices, known as measurement-DI-QRNG (MDI-QRNG). Unlike source-DI-QRNGs, which require well-characterized detectors~\cite{marangon2017source,cao2016source,avesani2018source,zhang2023realization}, MDI-QRNGs avoid the common situations where a practical detector deviates from the ideal detector model due to limited performance and unavoidable noise. Furthermore, the practicality of MDI-QRNG can be further improved by eliminating the need for single photon detection and designing protocols using room-temperature homodyne detectors \cite{wang2023provably}. Additionally, to provide a complete security analysis, it is essential to assume that the adversary possesses quantum side information. Compared to other semi-DI protocols based on MDI assumptions~\cite{nie2016experimental,cao2015loss,brask2017megahertz} and bounded dimension or energy assumptions~\cite{lunghi2015self,zhang2021simple,rusca2019self,avesani2021semi}, an MDI-QRNG protocol 
which accounts for quantum side information and employs homodyne detectors are able to provide high levels of security and practicality. This approach is fully compatible with photonic integration, which is necessary for miniaturized electronic devices.

Most chip-based QRNG demonstrations remain in the trusted-device category, based on vacuum fluctuation~\cite{bruynsteen2023100,bai202118,zheng20196} and qubit states collapse~\cite{jennewein2000fast,stanco2020efficient,zhang2023single}, with only a few demonstrating chip-based source-DI ~\cite{bertapelle2023high,kincaid2023source,du2023source} or energy-bound QRNGs~\cite{leone2020optical}. However, fully integrated chip-based semi-DI-QRNGs operating at room temperature have not yet been realized.

Various material platforms are available for chip-based QRNGs, including lithium niobate \cite{haylock2019multiplexed}, group III-V metals~\cite{abellan2016quantum}, perovskite \cite{argillander2023quantum}, and, notably, silicon \cite{prokhodtsov2020silicon,raffaelli2018generation}. The silicon platform stands out due to its high-speed modulators and detectors, allowing for the integration of all necessary components, including laser and electronic circuits \cite{atabaki2018integrating,wang2020integrated,luo2023recent}. Recent demonstrations in quantum computing \cite{zhang2021supercompact,vigliar2021error,qiang2021implementing} and quantum communication \cite{wei2020high,sibson2017integrated,zhang2019integrated,bunandar2018metropolitan,ma2016silicon} applications have underscored the capability of silicon photonic devices to encode and decode quantum states.
However, silicon has inherent limitations, particularly in linearity and phase-loss dependency on high-speed modulators and the efficiency and accuracy of photodiodes, which necessitates careful design considerations.    

In this work, we demonstrate the first self-testing QRNG chip with a fully integrated encoder and decoder, operating at room temperature using a homodyne detector. Our contributions to this work are two-fold:
(1) Theoretically, we develop a robust self-testing MDI-QRNG protocol that enhances the randomness expansion rate and tolerates more losses and noises on the chip. The protocol employs discrete modulated coherent states in quadrature phase shift keying (QPSK), combined with homodyne detection. Users can self-test randomness during operation by statistically verifying the detector's response against an ideal detector.
(2) Experimentally, we design a silicon photonic chip integrating all necessary components except the laser. Using 8-inch MPW processes and components, the fabricated chips approach the production level. The modulation scheme is designed to address the phase-loss dependency limitation.
The encoder uses an IQ modulator to avoid large phase shifts, and the basis selection phase modulator utilizes a Mazh-Zehnder modulator to compensate for the loss differences. The chip-based homodyne detector achieves a total efficiency of 69.1\%, and we obtain a secure randomness expansion rate of $5.11 \times 10^{-4}$ at a 10 MHz repetition rate. In each run, the experiment generated 15.33 Mbits of fresh randomness.

This work combines theoretical innovation with novel experimental schemes, demonstrating fully integrated silicon photonic chips for self-testing QRNG and providing a new level of security on the chip. Our chip paves the way towards the integration of practical QRNGs in systems that have small footprint and are secure.

\section{\label{sec:protocol}Self-testing quantum randomness expansion protocol}

Our randomness expansion protocol is based on a prepare and measure (P\&M) game $\mathcal{G}$. P\&M game $\mathcal{G}$ is an analogue of the more well-known non-local games; the latter has applications in Bell nonlocality~\cite{bell1964einstein, brunner2014bell} and device-independent protocols~\cite{primaatmaja2023security}. While non-local games are typically studied in entanglement-based scenarios where all the players are receivers of some quantum states, P\&M games typically involve a sender and a receiver. The game $\mathcal{G}$ can be characterised using the quantum state encoding, the probability distribution of the inputs for the sender and the receiver, and a scoring rule. 

A scoring rule assigns a score to each input-output configuration. In our proposed protocol, the score distribution is monitored to certify the amount of randomness generated. The scoring rule in this work can be seen in Eq.~\eqref{eq: scoring rule}.

\begin{figure*}[tbp]
\centering
\includegraphics[width=1.8\columnwidth]{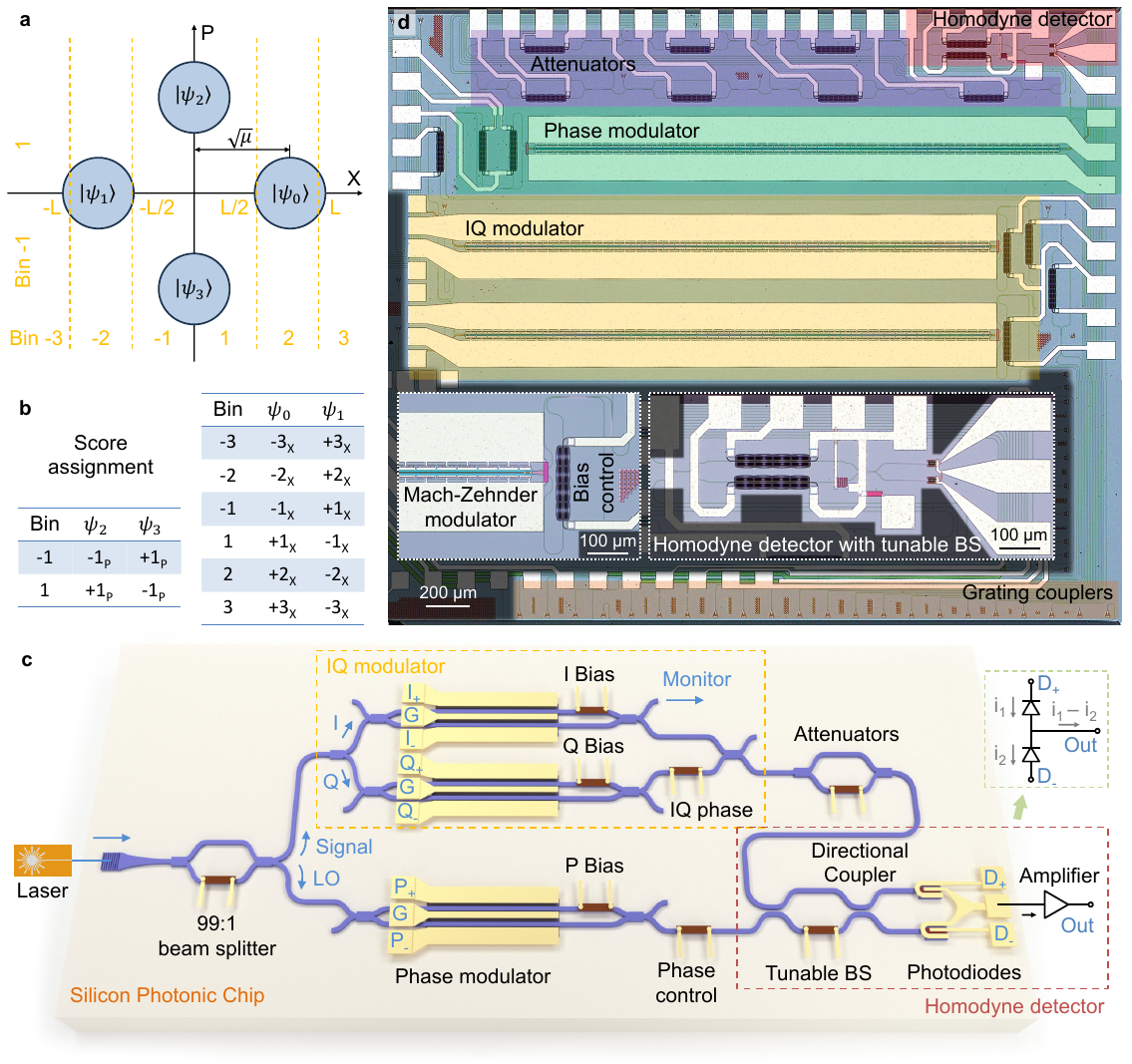}
\caption{Overview of the self-testing quantum randomness expansion chip. (a) Schematic of the encoding and decoding scheme. The input states are in quadrature phase shift keying (QPSK) with fixed intensity $\mu$. The receiver measures either X or P quadrature and discretizes the output into 6 or 2 bins, respectively. (b) The score of each round is assigned depending on the input states and the measured output bins. (c) Schematic of the quantum randomness expansion chip design. The laser is off-chip coupled and then split to signal and local oscillator (LO) path using a 99:1 beam splitter (BS). The LO light is modulated using a specially designed high-speed phase modulator with no phase-loss dependency. Another phase control compensates for all residue phase differences. The signal is encoded using an IQ modulator made of two parallel Mach-Zehnder modulators (MZM) and attenuated to the required intensity. The signal and LO are finally connected to a homodyne detector with tunable BS made of a directional coupler, which is used to fine-tune the detector balance. The photodiode output currents $i_1$ and $i_2$ are subtracted directly on the chip, and the output signal connects to a trans-impedance amplifier for detection. (d) A microscopic photo of the fabricated chip with false-colored key components. The Mach-Zehnder modulator with its bias controller and the homodyne detector is zoomed in and shown in the inset.}
\label{fig_overview}
\end{figure*}

Our proposed quantum random number generation protocol can be found below
\begin{enumerate}
    \item \textit{Input generation}. For each round $i\in [n]$, the round is randomly chosen between test round $(T_i = 1)$ and generation round $(T_i = 0)$ with test probability $\gamma$ and key probability $1-\gamma$. In each test round, the encoder, named Alice, selects her input as $X_i \in \mathcal{X} = \{0,1,2,3\}$, with an equal probability of 0.25 for each input. The receiver, named Bob, selects his input as $Y_i = 0$ when $X_i \in \{0,1\}$, and $Y_i = 1$ when $X_i \in \{2,3\}$. In each generation round, Alice and Bob fix their input as $X_i = 0$ and $Y_i = 1$. \\
    
    \item \textit{Quantum state preparation}. Alice prepares trusted quantum states $\ket{\psi_{X_i}} \in \{\ket{\sqrt{\mu}e^{ix\pi/2}}:x\in \mathcal{X}\}$ in the form of quadrature phase-shift keying (QPSK) with fixed intensity $\mu$ based on her input $X_i$, as shown in Fig.~\ref{fig_overview}a. 

    \item \textit{Quantum state measurement}. Bob measures the states prepared by Alice using an untrusted homodyne detector with his setting $Y_i\in \{0,1\}$, which corresponds to the local oscillator phase of 0 and $\pi/2$. The continuous output is discretized into $2m$ bins $\mathcal{B}=\{\pm 1,\pm 2,...,\pm m \}$ by dividing a finite interval $(-L,L)$ into equal $(2m-2)$ intervals, and the discretized result is recorded as $B_i$. In our current experiment, the number of bins is chosen as 6 and 2 when $Y = 0$ and 1, respectively.

    \item \textit{Parameter estimation}. For all generation rounds $(T_i = 0)$, we set $C_i = \perp$. On the other hand, for the test rounds $(T_i = 1)$, the scores $C_i = c$ are determined based on the following scoring rule,
    \begin{equation} \label{eq: scoring rule}
    c=  \begin{cases}
            +b_X,  & \text{$(X=0,B=+b$ or $X=1,B=-b)$}\\
            -b_X,  & \text{$(X=0,B=-b$ or $X=1,B=+b)$}\\
            +b_P,  & \text{$(X=2,B=+b$ or $X=3,B=-b)$}\\
            -b_P,  & \text{$(X=2,B=-b$ or $X=3,B=+b)$}\\
        \end{cases}.
    \end{equation}
    The score assignment in our current experiment is listed in Fig.~\ref{fig_overview}b. The frequencies $f(c)$ of each score $c$ from all $n$ rounds are calculated as
    \begin{equation}
    f(c) = \frac{1}{n} \sum^n_{i=1}\delta_{C_i,c} ,
    \end{equation}
    where $\delta_{j,k}$ denotes the Kronecker delta. The protocol is accepted if 
    \begin{equation}
    f(c) \leq \gamma(\omega_c + \delta_c), \qquad \forall c \neq \perp
    \end{equation}
    where $\vbf{\omega}$ is the ideal test score distribution, and $\vbf{\delta}$ is the tolerated score range, which means all the frequencies for $c \neq \perp$ should fall within a tolerated score distribution. Otherwise, the protocol is aborted. \\

\item \textit{Randomness extraction}. Our raw random string is formed by concatenating $\mathbf{R} = (\mathbf{B},\mathbf{T},\mathbf{X},\mathbf{Y})$. A quantum-proof strong extractor $\Ext$ with some uniformly random seed $\mathbf{S}$ is applied to the raw random string and outputs a $\ell$-bit string $\mathbf{Z} = \Ext(\mathbf{R}, \mathbf{S})$. The protocol finally outputs the concatenated string $\mathbf{K} = (\mathbf{Z}, \mathbf{S})$.

\end{enumerate}

Note that our randomness certification does not assume that the homodyne detector and quantum channel are characterised, nor do we assume that they behave independently or identically for each round. This provides our protocol with a practical advantage because detectors are typically hard to model and fully characterise. As long as the score distribution in the actual implementation closely resembles one that originates from a homodyne detection, the protocol can certify the randomness from the measurement outcomes. With fine-grained binning, the protocol can certify the measurement device better and tolerate more losses and noises. 

Moreover, we allow the adversary to have some pre-shared entanglement with Bob's untrusted detector before the protocol starts, i.e., the adversary possesses quantum side information, but we assume that the adversary does not update her quantum side information during the protocol operation. This assumption is not too restrictive because the quantum channel is not exposed to any external user during the protocol operation, and the whole chip is inside a secure location. 

Finally, the input randomness is recycled by hashing the random inputs together with the measurement outputs, which allows the protocol to consume more input randomness while maintaining a positive net random expansion rate. To be able to recycle the input randomness, the inputs are assumed to be isolated and are not leaked during and after the protocol. Also, the input is chosen randomly from a private and trusted source of randomness in an independent and identically distributed (i.i.d.) manner with no correlation with the adversary's side information.

As a result, the protocol is self-testing for the user. They only need to trust the quantum state preparation, while the performance of detectors and the generated randomness can be certified via the parameter estimation step during the operation. 

The security of our protocol is guaranteed by the quantum-proof strong extractor, such that either the protocol is aborted with high probability (e.g., in the case where the device does not behave as expected), or the output string is close to an ideal random string, which is uniformly distributed and independent from any quantum side-information and initial random seeds. 

Since our protocol consumes random bits during the input generation step, the net randomness expansion rate $\mathbf{r}_{\mathrm{net}} := (\ell - \ell_{\mathrm{in}})/n$ is the appropriate figure-of-merit for the protocol rather than the randomness generation rate. Here, $\ell_{\mathrm{in}}$ is the expected amount of randomness used during the protocol operation. Using the entropy accumulation theorem (EAT), our protocol has an expected randomness expansion rate of 
\begin{equation}
\label{eq_rate}
    \mathbf{r}_{\mathrm{net}} \gtrsim h - \mathcal{O}\left(\frac{1}{\sqrt{n}}\right), 
\end{equation} where $h$ depends on the tolerated score distribution and $\mathcal{O}(1/\sqrt{n})$ is a correction term for the finite-size effects, which scales with $1/\sqrt{n}$. The explicit expression for the correction term can be found in the Methods section (Section~\ref{subsec: randomness certification}).

\section{\label{sec:protocol}Silicon photonic chip design}

A silicon photonic chip is designed to implement our randomness expansion protocol. The chip schematic and a microscopic photo are shown in Fig.~\ref{fig_overview}c and d, respectively. The chip is fabricated by Advanced Micro Foundry Singapore using their active silicon-on-insulator (SOI) platform. The operation wavelength is chosen at O-band (1310.4~nm). Although, at this stage, we use an external laser and grating coupled to the chip, the full integration of laser and other control and supporting electronics is possible. The input laser is split into two paths for the signal and local oscillator (LO) using a Mach-Zehnder interferometer (MZI) as a tunable 99:1 beam splitter (BS). 

The LO path uses a high-speed carrier-depletion Mach-Zehnder modulator (MZM) with a specially designed driving scheme for phase selection between 0 and $\pi/2$. The centre common electrode is set as Ground (G), while the other two electrodes are supplied with two different driving signals $P_+$ and $P_-$. An additional thermo-optical phase shifter is included on one of the MZI arms for modulator bias and working point control. The main purpose of this design is to compensate for the phase-dependent loss caused by the carrier-depletion modulators. More detail is covered in Section \ref{sec:mod}. An additional thermo-optical phase shifter on the LO path compensates for all the phase differences between the signal and LO paths. 

The signal path uses an IQ modulator made of carrier-depletion MZMs to encode the required QPSK states. The IQ modulator is based on an MZI structure, where each arm contains a high-speed intensity modulator and modulates one of the I or Q quadratures. To form the IQ modulation, the modulated signals from both arms are combined with a fixed phase, set by the phase shifter named IQ phase. The intensity modulators in each arm are also based on MZMs. The driving signals are supplied with a push-pull configuration, which means that the centre electrode is grounded, while the top and bottom signals satisfy $I_+=-I_-$ and $Q_+=-Q_-$. Note here that we choose a very small driving voltage of $\pm$0.1~V, at which the phase-loss dependency is negligible. The bias controllers within MZMs modify the working point and ensure the output QPSK states have the same intensity. In Fig.~\ref{fig_overview}c, the port marked as Monitor is connected to an external power meter for fiber-chip alignment. The output from the IQ modulator is then connected to a series of attenuators made by MZI. The attenuators make sure the output signal is at the required intensity. 

Finally, both the signal and LO connect to the input ports of a homodyne detector. The homodyne detector uses a tunable 50:50 BS to mix the input signal and LO. Note that we use directional couplers rather than multimode interferometers (MMI) for their lower insertion loss. The BS outputs to two identical high-efficiency germanium photodiodes. A common electrode connects the anode and cathode of each photodiode, such that the common electrode outputs the subtracted photocurrent $i_1 - i_2$. The other electrodes $D_+$ and $D_-$ provide proper reverse bias for the photodiodes. The substrate photocurrent passes through a customised trans-impedance amplifier and is sampled using an oscilloscope. Detailed analysis for the homodyne detector is covered in Section \ref{sec:det}.

\section{\label{sec:mod}Phase-loss independent modulation}

The plasma dispersion effect, including carrier injection and carrier depletion, is the most common method to achieve high-speed modulation on silicon photonic platforms. However, the plasma dispersion effect modulator has a natural dependency between the output phase and loss. In our application, this dependency on the encoder side can distort the prepared states, while on the receiver side, the basis selection could be associated with intensity change, which can be a severe security loophole. In this section, the effect of phase-loss dependency is analyzed, and new solutions are introduced. 

\begin{figure}[tbp]
\centering
\includegraphics[width=1\columnwidth]{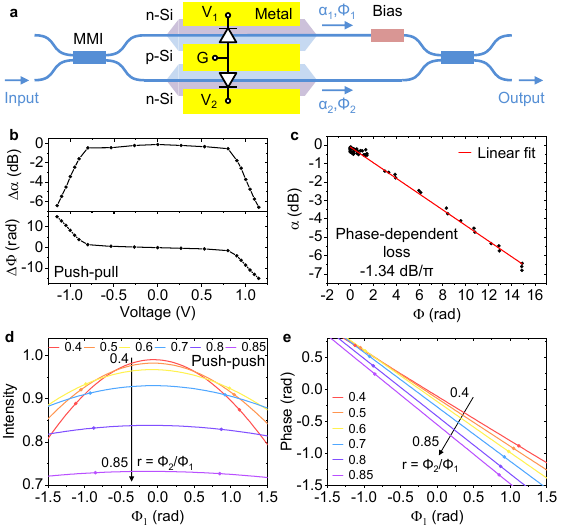}
\caption{Modulator phase-loss dependency analysis. (a) Schematic of the chip-based high-speed MZM, which contains two multi-mode interferometers (MMI) as beam splitters and one carrier depletion phase modulator on each arm. The centre common electrode is grounded while the other electrodes are supplied with different voltage $V$, resulting in a change in the loss $\alpha$ and phase $\Phi$ in each path. An additional phase shifter on one arm acts as bias control. (b) The measured loss and phase difference between the two arms as a function of supplied voltages, when $V_1 = -V_2$. (c) Extracted phase-loss dependency from (b), showing a linear relationship. (d,e) Simulated output intensity (d) and phase (e) from a push-push MZM under different modulated phases on each arm. The dots mark the working point where the output phase difference is $\pi/2$. 
}
\label{fig_mod}
\end{figure}

A schematic for the high-speed modulator is shown in Fig.~\ref{fig_mod}a. The modulator is based on an MZI structure, with carrier depletion phase modulators on both arms and their polarity reversed. The centre electrode is connected and grounded, while voltage $V_1$ and $V_2$ are applied on each arm. The output light after the modulator section gains phase $\Phi$ and loss $\alpha$. After an additional phase shifter for bias control, the lights are mixed on an MMI and output from one of the output ports. 

The phase-loss dependency is measured using a push-pull configuration, with $\Phi_1 = -\Phi_2$. The input wavelength is scanned across the whole O-band to acquire the extinction ratio and phase, which can further derive the loss. The difference in loss and phase as a function of the applied voltage is shown in Fig.~\ref{fig_mod}b. The fast-increasing zone above $\pm$0.6~V corresponds to one of the diodes turning on in the forward direction. The data can further derive the linear relationship between the phase and loss, as shown in Fig.~\ref{fig_mod}c. A phase modulator can introduce more than 35\% power difference in 0 to $1.5 \pi$ range, which proves the necessity of using an IQ modulator instead of a phase modulator for QPSK encoding. For our IQ modulator, the $\pm$ 0.1~V driving voltage corresponds to a phase shift of 0.04$\pi$ and a loss difference of maximum 0.05~dB, which can be compensated easily by the bias control. 

The requirement for the phase modulator on the LO path is different from the IQ modulator. The phase modulator requires a phase range of 0 to $\pi/2$ with minimum insertion loss to maintain the high efficiency of the homodyne detector. Here, we propose to operate the MZM in a push-push configuration, with $\Phi_2 = r\Phi_1$, where the nonlinearity of the MZM response compensates for the phase-dependent loss. Fig.~\ref{fig_mod}c and d show the simulated output intensity and phase of MZM as a function of modulated phase $\Phi$. All the curves use optimum bias values. The dots on each curve correspond to a phase range of $\pi/2$ with the same intensity. The result indicates that the output intensity fluctuates more and has a higher peak intensity when the ratio $r$ decreases. Considering a fixed $\pi/2$ phase range, there exists an optimum working point with $r=0.6$, where the output light has no phase-dependent loss and minimum total loss. 

\section{\label{sec:det}High-efficiency homodyne detector}

The homodyne detector is another key component of our protocol. Two identical waveguide-coupled germanium photodiodes together with an off-chip transimpedance amplifier are connected via wire bonds. Although the protocol does not require the detector to be characterised, we provide the specifications as a reference here to assist in demonstrating our protocol. 

\begin{figure}[tbp]
\centering
\includegraphics[width=1\columnwidth]{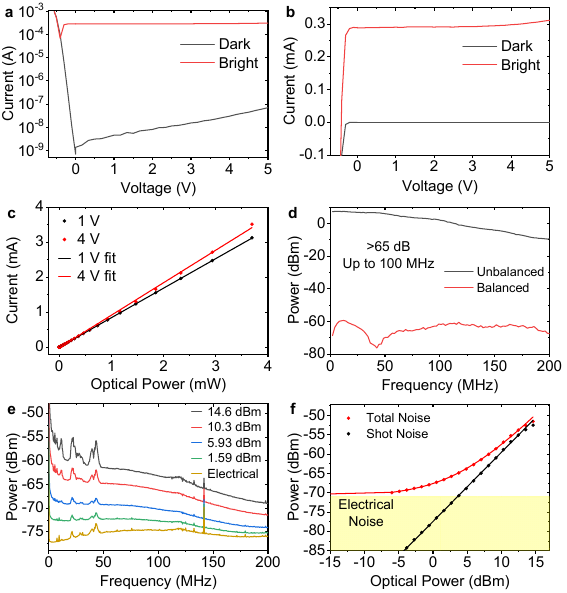}
\caption{Homodyne detector performance. (a,b) Current-Voltage (I-V) curve of photodiodes in (a) logarithmic scale and (b) linear scale, in both dark and bright conditions. (c) The photocurrent as a function of input optical power at different reverse bias voltages. (d) The detector balance test shows a common mode rejection ratio (CMRR) of $>$65 dB up to 100 MHz. (e) Shot-noise spectrum at different LO power, showing $>$11 dB shot-noise to electronic noise clearance up to 100 MHz. (f) Shot-noise as a function of LO power at 10 MHz clock frequency. $>$19 dB maximum clearance is observed.    
}
\label{fig_det}
\end{figure}
The quantum efficiency of photodiodes is measured first. The current-voltage (I-V) response of the photodiode is shown in Fig.~\ref{fig_det}a and b. Fig.~\ref{fig_det}a is shown in the logarithmic scale. The dark count is more than four orders of magnitude lower than photocurrents below 4~V, essential to reduce any electronic noise in the homodyne detector. Fig.~\ref{fig_det}b is shown in linear scale. An important feature to notice is the slight increase of photocurrent above 3~V, mainly due to impact ionization. Fig.~\ref{fig_det}c clearly shows that the responsivity of the photodiode at 4~V is higher than 1~V. However, in our application, quantum efficiency is defined as the effectiveness of converting incident photons into electrons. Because the impact ionization process happens after the photon-electron conversion process, thus the increase of photocurrent and responsivity can only be considered a gain rather than a quantum efficiency increase. As a result, the quantum efficiency should be measured at a low reverse bias voltage of 1~V and flat region on I-V curves, but the photodiodes can still be used at 4~V bias to provide a higher gain. We sample 3 different chips and get a quantum efficiency range of 76.9\% to 79.9\%. The lower bound of 76.9\% is used as the quantum efficiency value in our key rate calculation to maintain security. 

The photocurrent subtraction is achieved directly on the chip. The common mode rejection ratio (CMRR) is fine-tuned using an additional tunable beamsplitter made of MZI. The measured CMRR is $>$ 65 dB within the 100~MHz 3-dB bandwidth, as shown in Fig.~\ref{fig_det}d. The CMRR confirmed that the photodiode performances on the chip are similar, and the signal traces and wire bonds have minimum effect on the signal integrity. 

The measured shot noise spectrum is shown in Fig.~\ref{fig_det}e. Although the 3-dB bandwidth is about 100~MHz, the shot noise limited measurement can reach more than 500~MHz (not shown in the figure). Limited by the bandwidth of the homodyne detector, we choose a system repetition rate of 10 MHz, while the sample rate for the signal source and oscilloscope are kept at 1~GS/s. The shot noise as a function of local oscillator power at 10~MHz is shown in Fig.~\ref{fig_det}f. The power scale is different due to the difference in resolution bandwidth. The result shows that the homodyne detector saturates at about 13~dBm local oscillator power, with a shot noise clearance of 18.0~dB.  

Considering the insertion loss of the phase modulator on the LO path, the maximum shot-noise clearance is 14.2~dB, corresponding to a noise-equivalent efficiency $\eta_{eq} = 96.2\%$. Another loss we consider within the homodyne detector is the insertion loss of the tunable beam splitter, an MZI structure based on directional couplers. The total insertion loss is 0.30~dB = 93.3\%. As a reference, we also measured the insertion loss of a tunable beam splitter made of MMI as 0.72~dB = 84.7\%, which is too high for our protocol. The total homodyne efficiency, considering the quantum efficiency of photodiodes, the insertion loss of beamsplitter, and the noise equivalent efficiency, is $\eta = 69.1\%$. 

Moreover, we customised a set of new photonic chip components that can potentially provide better performance. The new photodiode at 1550 nm wavelength has a quantum efficiency of 92.4\%. Tunable beam splitter insertion loss is 0.30~dB = 93.3\%. Assume it can reach the same shot-noise clearance $\eta_{eq} = 96.2\%$. The total homodyne efficiency can reach 83.0\%.

\section{\label{sec:key}Randomness expansion simulation and experiment}

\begin{figure}[tbp]
\centering
\includegraphics[width=1\columnwidth]{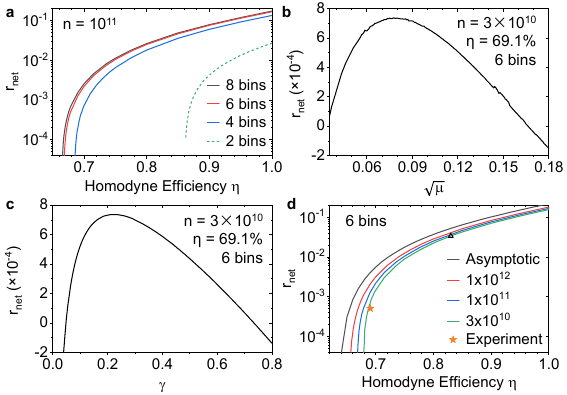}
\caption{Quantum randomness expansion rate simulation. (a) Rate as a function of homodyne detection efficiency with different numbers of discretization bins. Higher bin number shows better loss tolerance. The 2 bins result is reproduced from Ref. \cite{wang2023provably} (b,c) The effect of (b) input state amplitude and (c) test probability on the expansion rate. (d) Asymptotic rate and finite size rate at different numbers of rounds. The star marks our experimental result with a homodyne efficiency of 69.1\% and rounds number of $3\times 10^{10}$. The triangle marks the simulation point with a potentially upgraded homodyne efficiency of 83.0\%.
}
\label{fig_key}
\end{figure}

With the chip performance well characterised, the experimental parameters related to the randomness expansion protocol are optimized and listed in Table~\ref{tab_para}. Based on Eq.~\ref{eq_rate}, the simulated $\mathbf{r}_\mathrm{net}$ is shown in Fig.~\ref{fig_key}. Fig.~\ref{fig_key}a shows the effect of homodyne detection efficiency and the number of discretized bins on the homodyne detector output. As expected, more bin numbers increase the randomness expansion rate and possess more tolerance for detector efficiency. The data for 2 bins are reproduced from our recent work \cite{wang2023provably}, with a minimum acceptable efficiency of about 86\%, while the result for 4 to 8 bins shows a significant improvement at lower efficiency down to 67\%. Such improvement is essential considering the relatively higher loss and noise on photonic chips.    
\begin{table}[tbp]
\def\arraystretch{1.4}
    \caption{Experiment parameters}
    \centering
    \begin{tabular}{c c}
    \hline\hline 
        Homodyne efficiency $\eta$ & 69.1\% \\
        Number of bins $2m$ & 6 \\
        Number of rounds $n$ & $3\times10^{10}$ \\
        State amplitude $\sqrt{\mu}$ & 0.0672 \\
        Test probability $\gamma$ & 0.12 \\
        Completeness error $\epsilon_{com}$ & $1\times10^{-3}$ \\
        Soundness error $\epsilon_{sou}$ & $1\times10^{-6}$ \\
        Entropy accumulation error $\epsilon_{EA}$ & $1\times10^{-6}$ \\
        \;Randomness extraction parameter $\epsilon_{ext}$ \; & $1\times10^{-6}$ \\
        Smoothing parameter $\epsilon_{s}$ & \;$4.99\times10^{-7}$\; \\
    \hline\hline     
    \end{tabular}
    \label{tab_para}
\end{table}

Each point in Fig.~\ref{fig_key}a is optimized with respect to the state amplitude $\sqrt{\mu}$ and test probability $\gamma$, which are closely related to the experimental settings. Fig.~\ref{fig_key}b and c show the effect of these two parameters on the $\mathbf{r}_\mathrm{net}$. At the efficiency of our chip $\eta = 69.1\%$, the input state amplitude that provides a positive rate can vary from 0.035 to 0.165, while the test probability can vary from 0.05 to 0.73. In our experiment, we tried to keep the state amplitude optimum but used a slightly lower test probability to increase the tolerated score range $\delta$ and reduce the required accuracy of the experiment. 

Fig.~\ref{fig_key}d shows the finite size rate with 6 bins. Considering our chip-based homodyne efficiency of 69.1\%, we choose the number of rounds $n=3\times10^{10}$. The experimental result is marked as a star, while the triangle shows the simulation results using the customised chip components, with about two orders of magnitude improvement of rate. 

In our proof-of-principle experiment, the input generation step uses a pseudorandom input with a length of 100. Both the X and P quadrature for each state are measured to assist the system characterization. The unused measurements are discarded before the post-processing process. Considering the large number of rounds required, the system settings, including all the bias control and fiber-to-chip alignment, are monitored and calibrated constantly (every 100~s) to avoid any environmental drift. The data are also collected and digitally processed in blocks. The oscilloscope collects data with a memory size of 50~MSamples. After each calibration step, the oscilloscope memory is recorded 50 times to form a block, which corresponds to $2.5\times 10^7$ number of rounds. All the data within the $2.5\times 10^7$ rounds block are used to calculate the input state amplitude, whose accuracy is crucial in operating the protocol. We set a range of $\pm 0.0022$ for the acceptable amplitude. The whole data block is discarded if the amplitude exceeds the acceptable range. 

\begin{figure}[tbp]
\centering
\includegraphics[width=0.6\columnwidth]{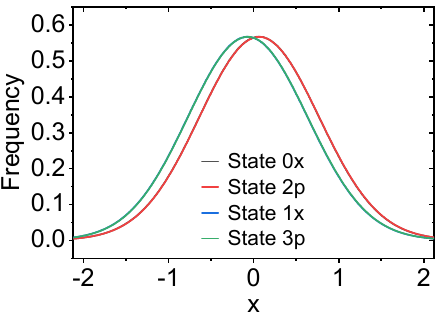}
\caption{Output quadrature distribution. The measured quadrature density distribution with all test states. The results clearly resemble Gaussian distributions. 
}
\label{fig_dist}
\end{figure}

The measured quadrature distribution for all four test states is shown in Fig.~\ref{fig_dist}, which faithfully resembles Gaussian distributions. Based on the scoring rule shown in Fig.~\ref{fig_overview}b, Table~\ref{tab_score} shows the target score $\omega$, tolerance $\delta$, and experiment-measured error. Thus, the data is accepted and can proceed to the post-processing step. The final secure random expansion rate is $5.11\times 10^{-4}$. With the number of rounds $n = 3\times 10^{10}$, the experiment generated fresh randomness in each round is 15.33 Mbits.

\begin{table}[tbp]
\def\arraystretch{1.4}
    \caption{Experimental score distribution}
    \centering
    \begin{tabular}{c c c c}
    \hline\hline 
        \;Score\; & $\omega$\; & $\delta$  & \;$f(c)/\gamma - \omega$\;  \\
    \hline
        $-3_X$ & 0.0659\; & 1.55$\times 10^{-5}$ & 0.61$\times 10^{-5}$  \\
        $-2_X$ & 0.0688\; & 1.58$\times 10^{-5}$ & -1.45$\times 10^{-5}$ \\
        $-1_X$ & 0.0931\; & 1.83$\times 10^{-5}$ & 0.11$\times 10^{-5}$  \\
        $-1_P$ & 0.2278\; & 2.85$\times 10^{-5}$ & -0.79$\times 10^{-5}$ \\
        $+3_X$ & 0.0927\; & 1.83$\times 10^{-5}$ & -0.36$\times 10^{-5}$ \\
        $+2_X$ & 0.0811\; & 1.71$\times 10^{-5}$ & 1.43$\times 10^{-5}$  \\
        1-all &  0.3706\; & 3.60$\times 10^{-5}$ & 0.46$\times 10^{-5}$ \\
    \hline\hline     
    \end{tabular}
    \label{tab_score}
\end{table}

\section{\label{sec:discussion}Discussion}

As a proof of principle experiment, the current required number of rounds, $n = 3\times 10^{10}$, still has room for improvement, primarily constrained by homodyne efficiency. Our new customised high-efficiency photodiode in the lab achieves 92.4\% efficiency at 1550 nm wavelength, providing a baseline for further enhancements. Shot-noise clearance can be improved through increased local oscillator power and improved amplifier. Recent demonstrations have showcased a clearance of 32.7 dB at 100 MHz \cite{milovanvcev2022chip}, highlighting the potential for improvement. 

Homodyme measurements offer the advantage of reaching higher system bandwidth at room temperature compared to single photon detectors. Considering chip integration, the active components, including modulators and detectors, are designed for classical communication with bandwidth exceeding 20 GHz. The homodyne detector with a bandwidth of up to 20 GHz has also been demonstrated \cite{ng2023240}, providing a substantial increase in the system clock rate by two to three orders of magnitude.   

A missing component for full chip integration is the on-chip laser, but it has already been demonstrated in major photonic foundries \cite{marinins2022wafer,li2022integrated}. The key parameters, such as a maximum power of 20~mW at 1550~nm and a linewidth below 1~MHz, are aligned closely with the table-top lasers we use now, indicating anticipated performances similar to those of this manuscript.   


Our current protocol involves a tradeoff between system complexity and the randomness expansion rate. Similar to our recent work \cite{wang2023provably}, increasing the number of bins and input states not only increases the number of statistics that need to be calculated and consumes more input randomness, but also imposes higher demands on component performance, such as the linearity of IQ modulator and the accuracy of homodyne detectors. Users can select the appropriate configuration based on their specific application requirements.

In conclusion, we propose and demonstrate a self-testing QRNG using a silicon photonic chip. The upgraded protocol, incorporating an uncharacterised measurement device, can tolerate detector efficiency down to 67\%. The silicon photonic chip integrates all necessary components, except the laser, on the same platform. The encoding uses an IQ modulator and a Mach-Zehnder modulator in a push-push configuration for detector base selection. Both modulators completely eliminate the effect of phase-loss dependency. The chip-based homodyne detector achieves a high total efficiency of 69.1\% at 1310 nm wavelength. The QRNG chip provides a total randomness expansion rate of $5.11 \times 10^{−4}$ at a repetition rate of 10 MHz. Our work paves the way for the application of quantum random numbers, providing provable security across diverse fields such as artificial intelligence, healthcare, energy services, blockchain, internet of things, and financial applications.

\section*{Acknowledgement}
The authors acknowledge funding support from the National Research Foundation of Singapore (NRF) Fellowship grant (NRFF11-2019-0001), NRF Quantum Engineering Programme 1.0 grant (QEP-P2 and QEP-P3), and MOE Tier 1 (A-8001168-00-00). 

\section*{Disclaimer}
This paper was prepared for informational purposes with contributions from the Global Technology Applied Research center of JPMorganChase. This paper is not a product of the Research Department of JPMorganChase. or its affiliates. Neither JPMorganChase. nor any of its affiliates makes any explicit or implied representation or warranty and none of them accept any liability in connection with this position paper, including, without limitation, with respect to the completeness, accuracy, or reliability of the information contained herein and the potential legal, compliance, tax, or accounting effects thereof. This document is not intended as investment research or investment advice, or as a recommendation, offer, or solicitation for the purchase or sale of any security, financial instrument, financial product or service, or to be used in any way for evaluating the merits of participating in any transaction.

\section*{Author contributions}
G.Z. and C.L. jointly conceived the idea. G.Z. and C.W. designed the photonic chips and experiments. I.W.P., K.T.G., and C.L. designed the randomness expansion protocol with the ramdomness certification proof. C.Y. and G.X. supported the photonic chip packaging. Z.G., C.W., S.Q.N., and H.J.N. conducted the experiments. All authors contributed to the discussion of experimental results. C.L. supervised and coordinated all the work. G.Z. and I.W.P. wrote the manuscript with contributions from all co-authors.

\section*{Competing interests}
The authors declare no competing financial interests.

\section*{Data availability}
All of the data that support the findings of this study are available in the main text. Source data are available from the corresponding author on request.

\section{Methods}
\subsection{Randomness certification} \label{subsec: randomness certification}
In this section, we shall certify the randomness that is generated by our protocol. First, we recall two criteria that a randomness generation protocol must satisfy, namely soundness and completeness.
\begin{enumerate}
    \item \textit{Soundness}\\
    Let $\Omega$ be the event in which the protocol is not aborted (i.e., it successfully produces a random output) and let $\Pr[\Omega]$ denotes the probability that this event occurs.\\
    For a given $\esound \in (0,1)$, a randomness generation protocol is said to be $\esound$-sound if the output $\vb{K}$ and the adversary's quantum side information $E$ satisfies
    \begin{equation} \label{eq: soundness}
        \Pr[\Omega] \cdot \frac{1}{2} \norm{\rho_{\vb{K}E | \Omega} - \tau_{L} \otimes \rho_{E|\Omega}}_1 \leq \esound,
    \end{equation}
    where $L$ denotes the length of the output $\vb{K}$ and $\tau_{L}$ is the maximally mixed state with dimension $2^L$. $\rho_{\vb{K}E|\Omega}$ denotes the state of the output of an implementation of the protocol which is not aborted.

    \item \textit{Completeness}\\
    Let $\ecom \in (0,1)$ be given. A randomness generation protocol is said to be $\ecom$-complete if
    \begin{equation}
        \Pr[\Omega]_{\mathrm{honest}} \geq 1 - \ecom.
    \end{equation}
    Here, the subscript ``honest'' indicates that the probability of the protocol not being aborted is evaluated assuming that the device behaves honestly (i.e., the device behaves as expected).
\end{enumerate}

The soundness criterion formalizes the intuitive goal that a randomness generation protocol should produce an output that is uniform and independent of the adversary's side information. However, in a self-testing randomness generation protocol, we must allow the protocol to be aborted when the protocol detects that the output does not have sufficient entropy (e.g., due to some malfunction in the measurement device). Therefore, the soundness criterion demands that either the protocol aborts with high probability or the output is close to being uniform and independent from any side information.

The completeness criterion formalizes the desired property that when the device is behaving honestly, the probability of aborting the protocol should be small. When the device is behaving honestly, the protocol can still be aborted due to statistical fluctuations in the parameter estimation step. In our case, the honest behaviour is described by some homodyne measurement which performs X and P quadrature measurements with known detection efficiency.

Before we present the randomness certification, we shall explicitly state the assumptions that we make to certify the generated randomness. Throughout this work, we assume the following:
\begin{enumerate}
    \item Quantum theory is correct
    \item The device consist of a trusted and characterised source of quantum states (which we personify as Alice), an uncharacterised measurement device (which we personify as Bob) and a trusted classical processing unit which supplies random inputs, stores measurement outcomes and performs the necessary post-processing as specified by the protocol.
    \item The device is located in a secure location and is sufficiently isolated from the environment during the execution of the protocol. We allow the device to share some correlation (which can be classical or quantum) with the adversary prior to the protocol execution.
    \item The device is equipped with some trusted random seed that is uncorrelated to the adversary's side information, nor to the devices (i.e., the source and the measurement device) themselves.
\end{enumerate}

The proof for the protocol's completeness is considerably more straightforward. When the device's honest behaviour is characterised, the expected score distribution is known. In this case, the protocol will only abort due to statistical fluctuations, and hence to ensure completeness, we only need to set $\vbf{\delta}$ to be large enough to allow for these fluctuations.

Let us denote the expected score distribution for the honest behaviour as $\vbf{p}^{\mathrm{hon}} := (1-\gamma, \gamma \vbf{\omega}^{\mathrm{hon}})$. Here, the first entry refers to the probability of obtaining $C = \perp$, whereas the $\vbf{\omega}^{\mathrm{hon}}$ gives the conditional score distribution given that the round is a test round. Denote the range of $C$ as $\mathcal{C}$ and we write $\mathcal{C}' = \mathcal{C} \setminus \{\perp\}$ as the range of $C$ for the test rounds. Our strategy to prove the completeness of the protocol is by deriving an upper bound
\begin{equation} \label{eq: eps complete c}
    \Pr[f(c) > \gamma(\omega^{\mathrm{hon}}_c + \delta_c)] \leq \varepsilon_{\mathrm{com}, c}
\end{equation}
for each $c \in \mathcal{C}'$. Then, the union bound implies that
\begin{equation} \label{eq: union bound}
    \ecom \leq \sum_{c \in \mathcal{C}'} \varepsilon_{\mathrm{com}, c}.
\end{equation}

In this work, to obtain the upper bound $\varepsilon_{\mathrm{com},c}$ in Eq.~\eqref{eq: eps complete c}, we use the following strategy. For each $c \in \mathcal{C}'$, we consider a sequence of random variables that indicate whether for a given round $i$, Alice and Bob obtains the score $C_i = c$. For each $i$, this random variable is a Bernoulli random variable, and since the honest behaviour between rounds will be independent, the number of rounds with score $c$ follows a binomial distribution, with $n$ trials and probability $\gamma \omega^{\mathrm{hon}}_c$. We can then use the bounds on binomial distribution presented in Ref.~\cite{zubkov2013complete} to obtain the upper bounds
\begin{equation}
    \varepsilon_{\mathrm{com}, c} = 1 - F\left(n, \gamma \omega^{\mathrm{hon}}_c, \floor{n\gamma(\omega^{\mathrm{hon}}_c + \delta_c)} \right),
\end{equation}
where
\begin{align}
    D(q,p) &:= q \ln \left(\frac{q}{p}\right) + (1-q) \ln \left(\frac{1-q}{1-p}\right),\\
    \Phi(a) &:= \frac{1}{\sqrt{2\pi}} \int_{-\infty}^{a} \mathrm{d}x \, \exp \left(-\frac{x^2}{2}\right),\\
    F(n,p,k) &:= \Phi \left(\mathrm{sign}\left(\frac{k}{n} - p\right) \sqrt{2 n D\left(\frac{k}{n},p \right)}\right).
\end{align}
The union bound~\eqref{eq: union bound} concludes the completeness proof.

For soundness, assuming that the adversary's side information is limited by quantum theory, we can invoke leftover hashing against quantum side information~\cite{tomamichel2011leftover}. The leftover hash lemma states that we can use a two-universal family of hash functions as a quantum-proof strong extractor $\Ext$. If $\vb{R}$ is the input of the extractor and $\vb{S}$ is a uniformly chosen random seed, and we let $\vb{Z} = \Ext(\vb{R}, \vb{S})$, then the output of the randomness extraction step satisfies the following
\begin{multline} \label{eq: leftover hash lemma}
    \frac{1}{2} \norm{\rho_{\vb{Z} \vb{S} E} - \tau_{\ell} \otimes \tau_{s} \otimes \rho_E}_1 \\
    \leq 2 \esmooth + 2^{-\frac{1}{2}\left( H_{\min}^{\esmooth}(\vb{R}|E)_{\rho} - \ell + 2\right)},
\end{multline}
for any $\esmooth \in (0,\sqrt{\Tr[\rho]})$. Here, $\ell$ is the length of the output $\vb{Z}$ while $s$ is the length of the seed $\vb{S}$ that we require for randomness extraction. Thus, the concatenated length $L = \ell + s$.

If we set the right hand side of the inequality \eqref{eq: leftover hash lemma} to $\esound$, then it would be sufficient to prove a lower bound on the conditional smooth min entropy $H_{\min}^{\esmooth}(\vb{R}|E)_{\rho_{\vb{R}E}}$ to establish the soundness of the protocol. However, in practice, it is not easy to establish a lower bound on the conditional smooth min-entropy directly. Instead, we consider two cases (that are non-exhaustive):
\begin{itemize}
    \item[(a)] $\Pr[\Omega] < \eEAT$ for some fixed $\eEAT \in (0,1)$
    \item[(b)] $\Pr[\Omega] \geq \eEAT$ and $H_{\min}^{\esmooth}(\vb{R}|E)_{\rho_{\vb{R}E \wedge \Omega} } \geq k$ for some constant $k$ that we will determine later.
\end{itemize}
Here, $\rho_{\vb{R}E \wedge \Omega} = \Pr[\Omega] \cdot \rho_{\vb{R}E | \Omega}$ is a sub-normalised state associated to the case where the protocol is not aborted.

For case~(a), we can use $\esound = \eEAT$ as an upper bound for the soundness criterion~\eqref{eq: soundness}. For case~(b), we can use $k$ as a lower bound the conditional smooth min-entropy and invoke the leftover hash lemma. If we set the exponential term of Eq.~\eqref{eq: leftover hash lemma} to be equal to $\eExt$, then we can set $\esound = 2 \esmooth + \eExt$ and the maximum output length that we can have is given by
\begin{equation}
    \ell = \floor{k - 2\log_2 \left(\frac{1}{\eExt}\right) + 2}
\end{equation}
Combining the two cases, we conclude that the soundness parameter can be set as
\begin{equation}
    \esound = \max \{\eEAT, 2 \esmooth + \eExt \}.
\end{equation}

Now, our task is to find the constant $k$, which is a lower bound on the conditional smooth min-entropy in case (b). Recall that in our protocol, the input $\vb{R}$ to the randomness extractor $\Ext$ is given by the concatenation of the measurement outcomes $\vb{B}$ and the inputs $(\vb{T}, \vb{X}, \vb{Y})$. The conditional smooth min-entropy can be lower bounded using the chain rule for smooth entropies~\cite{vitanov2013chain}
\begin{multline} \label{eq: chain rule for smooth entropies}
    H_{\min}^{\esmooth}(\vb{R}|E)_{\rho_{\wedge \Omega}}  \\\geq H_{\min}^{\epsilon_1}(\vb{B}|\vb{T} \vb{X} \vb{Y}, E)_{\rho_{\wedge \Omega}}
    + H_{\min}^{\epsilon_2}(\vb{T}\vb{X}\vb{Y}|E)_{\rho_{\wedge \Omega}} \\- \log_2\left(\frac{2}{\esmooth - \epsilon_2 - 2\epsilon_1} \right).
\end{multline}

Lower bounding the second term is easy since we have assumed that the inputs are chosen from a trusted random seed which is also independent of the adversary's side information $E$. Furthermore, the inputs are identical and independently distributed (IID), which allows us to invoke the quantum asymptotic equipartition property (QAEP)~\cite{tomamichel2009fully}. The QAEP~\footnote{We use the version of QAEP given in Corollary 4.10 of Ref.~\cite{dupuis2020entropy}} states that
\begin{equation}
    H_{\min}^{\epsilon_2}(\vb{T}\vb{X}\vb{Y}|E)_{\rho} \geq n H(T_i, X_i, Y_i|E) - \xi \sqrt{n},
\end{equation}
where 
\begin{equation}
    \xi = 2 \log_2(1 + 4m)\sqrt{1 - 2 \log_2 \epsilon_2}.
\end{equation}

From our assumption that the inputs are independent from $E$, we have
\begin{align}
    H(T_i, X_i, Y_i | E) &= H(T_i, X_i, Y_i) \nonumber\\
    &= H(T_i) + H(X_i, Y_i|T_i) \nonumber\\
    &= H(T_i) + \gamma H(X_i, Y_i| T_i = 1) \nonumber\\
    &= h_2(\gamma) + 2 \gamma,
\end{align}
where in the first line, we use the independence from $E$, in the second line, we use the chain rule for Shannon entropy, in the third line, we decompose the conditional entropy term and drop the $H(X_i, Y_i|T_i = 0)$ term since the inputs are fixed in the generation rounds. In the last line, we compute the value of each term, where $h_2$ denotes the binary entropy function.

While we have obtained the conditional smooth min-entropy $H_{\min}^{\epsilon_2}(\vb{TXY}|E)_{\rho}$ evaluated on the unconditioned state $\rho_{\vb{TXY}E}$, we can convert this to the state $\rho_{\vb{TXY}E \wedge \Omega}$ using Lemma 10 of Ref.~\cite{tomamichel2017largely} as long as $\epsilon_2 \in (0, \sqrt{\Pr[\Omega]})$. This can be done by choosing $\epsilon_2 < \eEAT < \sqrt{\eEAT} \leq \sqrt{\Pr[\Omega]}$ since we are in case (b). Thus, we have established the bound
\begin{equation}
    H_{\min}^{\epsilon_2}(\vb{TXY}|E)_{\rho_{\wedge \Omega}} \geq n \left[h_2(\gamma) + 2 \gamma \right] - \xi \sqrt{n}.
\end{equation}

For the first term of \eqref{eq: chain rule for smooth entropies}, we can use the entropy accumulation theorem (EAT)~\cite{dupuis2019entropy, dupuis2020entropy} to bound $H_{\min}^{\epsilon_1}(\vb{B}|\vb{T}\vb{X}\vb{Y},E)_{\rho_{ \wedge \Omega}}$ in case~(b). EAT relates the conditional smooth min-entropy for the entire protocol (i.e., accumulated over $n$ rounds) to a single-round quantity called the min-tradeoff function $f_{\vbf{\nu}}$ (for the precise definition, we refer the readers to Refs.~\cite{dupuis2019entropy, dupuis2020entropy, brown2019framework}). This makes the analysis tractable as the min-tradeoff function can be constructed using semidefinite programming (SDP) method~\cite{wang2019characterising, brown2019framework, wang2023provably}.

There is some condition to the application of EAT, namely that the inputs and outputs satisfy the so-called Markov chain condition~\cite{dupuis2019entropy, dupuis2020entropy}. Based on the same argument as the one presented in our previous experiment~\cite{wang2023provably}, this is satisfied in our experiment as the inputs are chosen from an independent and trusted random seed and the device is sufficiently isolated such that the adversary's quantum side information is not updated in the protocol.
 
To construct the min-tradeoff function, let $\pg(\vbf{\omega})$ denotes the single-round guessing probability subject to the score constraints $\vbf{\omega}$. Formally, it is defined as the solution to the following optimization
\begin{equation} \label{eq: primal}
    \begin{split}
        \max_{\{M_{b|y}\}, \{E_e\}} \,&\sum_{b = 0}^1 \bra{\varphi_0} M_{b|1} \otimes E_{b} \ket{\varphi_0}\\
        \mathrm{s.t.} \quad & \sum_{b,x,y} \kappa^{(c)}_{b,x,y} \bra{\varphi_x} M_{b|y} \otimes \mathbb{I} \ket{\varphi_x} = \omega_c,
    \end{split}
\end{equation}
where $\ket{\varphi_x}$ is the state shared by Bob and the adversary (which includes any pre-shared entanglement between them). Since we assume the correctness of quantum theory, we can describe measurements using positive operator-valued measure (POVM). Here, $\{M_{b|y}\}_{b,y}$ is the POVM elements describing Bob's untrusted measurement, $\{E_e\}_e$ is the POVM elements describing the adversary's measurement. $\kappa_{b,x,y}^{(c)}$ are the coefficients that gives the appropriate weight to the game's scores.

Let $\alpha_{\vbf{\nu}}$ and $\vbf{\lambda}_{\vbf{\nu}}$ be constants such that
\begin{equation}
    \pg(\vbf{\omega}) \leq \alpha_{\vbf{\nu}} + \vbf{\lambda}_{\vbf{\nu}} \cdot \vbf{\omega}
\end{equation}
for any score distribution $\vbf{\omega}$ that is compatible with quantum theory. The constants $\alpha_{\vbf{\nu}}$ and $\vbf{\lambda}_{\vbf{\nu}}$ can be found by taking the the dual of optimization~\eqref{eq: primal} for $\pg(\vbf{\nu})$ (which motivates the subscripts on $\alpha_{\vbf{\nu}}$ and $\vbf{\lambda_{\nu}}$).

Following the derivation of Theorem 1 presented in Ref.~\cite{wang2023provably}, EAT implies that the conditional smooth min-entropy $H^{\epsilon_1}_{\min}(\vb{B}|\vb{T}\vb{X}\vb{Y},E)_{\rho_{|\Omega}}$ is given by
\begin{multline}
    H^{\epsilon_1}_{\min}(\vb{B}|\vb{T}\vb{X}\vb{Y},E)_{\rho_{|\Omega}} \\
    \geq n h  - n \left[\beta V + \beta^2 K \right] - \frac{1}{\beta} [1 - 2\log_2 (\eEAT \epsilon_1)],
\end{multline}
where
\begin{align}
    h &= 2(1-\gamma) \left(1 - \alpha_{\vbf{\nu}} - \vbf{\lambda}_{\vbf{\nu}} \cdot \tilde{\vbf{\omega}}\right),\\
    V &= \frac{\ln 2}{2} \Bigg(\log_2(4m + 1) \nonumber \\
    &\hspace{1.5cm}+ \sqrt{2 + \frac{4(1-\gamma)^2(\lambda_{\max} - \lambda_{\min})^2}{\gamma}} \Bigg)^2,\\
    K &= \frac{2^{\beta(\log_2 2m +2(1-\gamma)(\lambda_{\max} - \lambda_{\min}))}}{6 \ln 2 (1-\beta)^3} \nonumber \\
    & \qquad \times \ln^3\left(2^{\log_2 2m + 2(1-\gamma)(\lambda_{\max} - \lambda_{\min})} + e^2\right),
\end{align}
where $\lambda_{\max} = \max_c \vbf{\lambda_\nu}$ and $\lambda_{\min} = \min_c \vbf{\lambda_\nu}$. Letting $c_{\min} = \mathrm{argmin}_c \vbf{\lambda_{\nu}}$, we define $\tilde{\vbf{\omega}}$ as
\begin{equation}
    \tilde{\omega}_c = 
    \begin{cases}
        \omega_c + \delta_c & \text{if $c \neq c_{\min}, \perp$} \\
        \omega_c - \sum_{c \neq c_{\min}, \perp} \delta_c & \text{if $c = c_{\min}$}.
    \end{cases}
\end{equation}

Thus, we have a lower bound on the conditional smooth min-entropy $H^{\epsilon_1}_{\min}(\vb{B}|\vb{TXY},E)_{\rho_{|\Omega}}$ evaluated on the state $\rho_{\vb{BTXY}E|\Omega}$. However, we are interested in the conditional smooth min-entropy evaluated on the \textit{sub-normalised} version of the state $\rho_{\vb{BTXY}E \wedge \Omega}$. We shall show that we indeed have $H^{\epsilon_1}_{\min}(\vb{B}|\vb{TXY},E)_{\rho_{\wedge \Omega}} \geq H^{\epsilon_1}_{\min}(\vb{B}|\vb{TXY},E)_{\rho_{|\Omega}}$.

First, recall that given the state $\rho_{AB}$, the conditional min-entropy can be defined as~\cite{tomamichel2010duality}
\begin{equation}
    H_{\min}(A|B)_{\rho} = \max_{\sigma_B} \sup \{\zeta: \rho_{AB} \preceq 2^{-\zeta} \mathbb{I} \otimes \sigma_B \}.
\end{equation}
Meanwhile, conditional smooth min-entropy is defined as~\cite{tomamichel2010duality}
\begin{equation}
    H_{\min}^{\varepsilon}(A|B)_{\rho} = \max_{\hat{\rho}: \, \mathcal{P}(\rho, \hat{\rho} )\leq \varepsilon} H_{\min}(A|B)_{\hat{\rho}},
\end{equation}
where $\mathcal{P}(\rho, \hat{\rho})$ denotes the purified distance between the state $\rho$ and $\hat{\rho}$. considering the above two definitions, if $H_{\min}^{\varepsilon}(A|B)_{\rho} = h_*$, then there exists a state $\hat{\rho}_{AB}$ and state $\sigma_B$ such that
\begin{equation}
    \hat{\rho}_{AB} \preceq 2^{-h_*} \mathbb{I}_A \otimes \sigma_B
\end{equation}
and
\begin{equation}
    \mathcal{P}(\rho_{AB}, \hat{\rho}_{AB}) \leq \varepsilon.
\end{equation}
Now, let $0 < p \leq 1$. Let $\rho' = p \rho$ and $\hat{\rho}' = p \hat{\rho}$. Clearly, multiplying each state by $p$ is a completely positive, trace non-increasing (CPTNI) map. Due to monotonicity of purified distance under CPTNI maps, we have
\begin{equation}
    \mathcal{P}(\rho'_{AB}, \hat{\rho}'_{AB}) \leq  \mathcal{P}(\rho_{AB}, \hat{\rho}_{AB}) \leq \varepsilon.
\end{equation}
Furthermore, we have
\begin{equation}
    \hat{\rho}'_{AB} \preceq \hat{\rho}_{AB} \preceq 2^{-h_*} \mathbb{I}_A \otimes \sigma_B.
\end{equation}
This implies that $H_{\min}^{\varepsilon}(A|B)_{\rho'} \geq h_*$ as well. In other words, multiplying a state by a number $0 < p \leq 1$ does not reduce the conditional smooth min-entropy. We apply this property to our case where the initial state is $\rho_{\vb{BTXY}E|\Omega}$ and the final state is $\rho_{\vb{BTXY}E \wedge \Omega} = \Pr[\Omega] \cdot \rho_{\vb{BTXY}E|\Omega}$.

Therefore, we have lower bounds for both $H_{\min}^{\epsilon_1}(\vb{B}|\vb{TXY},E)_{\rho_{\wedge \Omega}}$ and $H_{\min}^{\epsilon_2}(\vb{TXY}|E)_{\rho_{\wedge \Omega}}$. We can then use the chain rule \eqref{eq: chain rule for smooth entropies} to derive the lower bound $k$ for case (b)
\begin{multline}
    k = n[h + h_2(\gamma) + 2 \gamma] \\
    - n(\beta V + \beta^2 K) - \frac{1}{\beta} [1 -2\log_2(\eEAT \epsilon_1)] \\
    - \xi \sqrt{n} - \log_2\left(\frac{2}{\esmooth - \epsilon_2 - 2 \epsilon_1}\right)
\end{multline}
This concludes the soundness proof of the protocol.

Now, to calculate the expected randomness expansion rate, we have to consider the amount of randomness that we need to generate the inputs. There are some efficient algorithms (such as the interval algorithm~\cite{hoshi1997interval}) that convert a uniform bit strings into biased bit strings of shorter length, which we need for our protocol. For the interval algorithm, the required input bit string has length of
\begin{equation}
    \ell_{\mathrm{in}} = n H(T_i, X_i, Y_i) + 3 = n [h_2(\gamma) + 2 \gamma] + 3.
\end{equation}

We compare this to the output length
\begin{multline}
    \ell = \Bigg \lfloor n[h + h_2(\gamma) + 2 \gamma] \\
    - n(\beta V + \beta^2 K) - \frac{1}{\beta} [1 - 2 \log_2(\eEAT \epsilon_1)] - \xi \sqrt{n} \\
    - \log_2\left(\frac{2}{\esmooth - \epsilon_2 - 2 \epsilon_1}\right) - 2 \log_2 \left(\frac{1}{\eExt}\right) + 2\Bigg\rfloor.
\end{multline}
This gives the expected randomness expansion rate
\begin{multline} \label{eq: actual rate}
    \vb{r}_{\mathrm{net}}
    = h - \beta (V + \beta K) - \frac{1 - 2 \log_2(\eEAT \epsilon_1)}{n \beta} - \frac{\xi}{\sqrt{n}} \\
     - \frac{\log_2\left(\frac{2}{\esmooth - \epsilon_2 - 2 \epsilon_1}\right) + 2 \log_2\left(\frac{1}{\eExt}\right) - 1}{n}.
\end{multline}
To obtain the scaling that we claimed in Eq.~\eqref{eq_rate}, we can choose $\beta \sim \mathcal{O}(1/\sqrt{n})$. However, in our plots, we optimize $\beta$ 
 in Eq.~\eqref{eq: actual rate} to maximize the expected randomness expansion rate.

\subsection{Experimental Setup} \label{subsec: experiment}
The photonic chip is wire bond packaged on a customised PCB that hosts all the electrical input-output ports. The direct current (DC) ports are controlled by an off-chip multi-channel power supply (Qontrol Q8iv). The radio frequency (RF) ports are connected to arbitrary waveform generators (Zurich Instrument HDAWG4). The homodyne detector amplifier is customised based on the Texas Instrument OPA847 operational amplifier chip. The photonic chip is put close to the amplifier chip to minimize any input parasitic. The output signal from the amplifier is collected using an oscilloscope (Lecory WavePro HD) and analyzed on a computer.

The optical input is an O-band tunable laser (EXFO T100S) with a maximum output power of 14.5~mW. The laser passes a polarization controller and is coupled to the photonic chip with the assistance of a precision alignment stage. The output light splits into two paths. One part is measured using an off-chip power meter (Thorlab PM100D), and the other connects to a photodetector for waveform analysis. 

During the experiment process, all the heaters on the chip are calibrated periodically to avoid any environmental drift. The homodyne tunable BS is fine-tuned first with the signal path set to a high attenuation state and the LO path with an arbitrary modulation. The modulation on LO is minimized when the CMRR is at the optimum state. Then, the LO phase modulator P bias is fine-tuned to make sure the intensity on the X and P quadratures are the same. The IQ modulator biases are calibrated next to ensure accurate QPSK states. Finally, the attenuators are used to set the state intensity. The whole calibration process is repeated every 100~s (approximately), which is the time for 50 rounds of oscilloscope data collection. The frequent calibration process ensures the high requirement of state intensity.    

\bibliography{apssamp}

\end{document}